\documentclass[aps,prd,onecolumn,amssymb]{revtex4}
\usepackage{graphicx,bm}
\usepackage[english]{babel}
\usepackage{amsmath}
\usepackage{amsfonts}
\begin{document}

\newcommand{\vv}[1]{\vec{\w{#1}}}
\newcommand{\vn}[1]{\vec{#1}}
\newcommand{\vp}[1]{\overrightarrow{#1}}
\newcommand{\vpp}[4]{\overrightarrow{{#1}#2{#3}#4}}
\newcommand{\uu}[1]{\underline{\w{#1}}}
\newcommand{\uup}[1]{\underline{#1}}
\newcommand{\upp}[4]{\underline{{#1}#2{#3}#4}}
\newcommand{\encadre}[1]{\fbox{$\displaystyle #1$}}
\newcommand{\der}[2]{\frac{\partial #1}{\partial #2}}
\newcommand{\dder}[3]{\frac{\partial^2 #1}{\partial #2\partial #3}}
\newcommand{\dderr}[2]{\frac{\partial^2 #1}{\partial {#2}^2}}
\newcommand{\dert}[2]{{\partial #1}/{\partial #2}}
\newcommand{\dderrt}[2]{{\partial^2 #1}/{\partial #2}^2}
\newcommand{\w}[1]{\bm{#1}}

\newcommand{\Lie}[1]{\mathcal{L}_{#1}\,}
\newcommand{\Liec}[1]{{\mathcal{L}}_{\vv{#1}}\,}

\newcommand{\be}{\begin{equation}}
\newcommand{\ee}{\end{equation}}
\newcommand{\bea}{\begin{eqnarray}}
\newcommand{\eea}{\end{eqnarray}}
\newcommand{\ben}{\begin{enumerate}}
\newcommand{\een}{\end{enumerate}}
\newcommand{\nn}{\nonumber \\}
\newcommand{\e}{\mathrm{e}}

\title{Neutron Stars in $f(R)$-Gravity and Its Extension with a Scalar Axion Field}
\author{Artyom V. Astashenok$^{1,}$\footnote{E-mail: aastashenok@kantiana.ru},
Sergei D. Odintsov$^{2,3,4,}$\footnote{E-mail:
odintsov@ieec.uab.es}}

\affiliation{$^{1}$Immanuel Kant Baltic Federal University, Nevskogo str., 14, 236041 Kaliningrad, Russia \\
$^{2}$ICREA, Passeig Luis Companys, 23, 08010 Barcelona, Spain \\
$^3$Institute of Space Sciences (IEEC-CSIC) C. Can Magrans
s/n, 08193 Barcelona, Spain\\
$^4$International Laboratory of Theoretical Cosmology, TUSUR,
634050 Tomsk, Russia}

\begin{abstract}
We present a brief review of general results about non-rotating
neutron stars in simple  $R^2$ gravity and its extension with a
scalar axion field. Modified Einstein equations are presented for
metrics in isotropical coordinates. The mass--radius relation,
mass profile and dependence of mass from central density on
various equations of state are given in comparison to general
relativity.
\end{abstract}

\maketitle

\section{Introduction}

Neutron stars are very interesting objects for the possible
verification of not only physical models of dense matter but
various theories of modified gravity. The~most simple $R^2$
gravity and its possible extensions were investigated  as a
possible alternative to general relativity in many papers (see,
for example, ~\citep{Capozz2011,Capozz2012}). The~main motivation
comes from cosmology with the discovery of the accelerated
expansion of universe \citep{Riess-1,Perlmutter,Riess-2}.

According to the standard approach, this acceleration occurs due
to nonzero vacuum energy consisting of nearly 70\% of the global
energy budget of the universe. The~remaining 28\%, clustered in
galaxies and clusters of galaxies, consists of baryons (only 4\%)
and cold dark matter (CDM), the nature of which is unclear.
Another paradigm is the description of cosmological acceleration
in frames of modified gravity
\citep{Capozziello1,Odintsov1,Turner}. It is interesting to note
that a unified description of cosmological evolution, including
epochs of matter and radiation dominance, is possible in the
$f(R)$ theory \citep{Nojiri,Nojiri-5,Capoz,Olmo-2,Cruz,Nojiri-4}.

However, in context of modification of general relativity, one need consider not only the cosmological level, but stellar structures too, especially compact relativistic objects (neutron stars and black holes). The~possible deviations from GR 
can be detected due to the extremely strong gravitational field in
the centers of relativistic~stars.

This paper presents a brief review of general results about
non-rotating neutron stars in simple  $R^2$ gravity and its
extension with a scalar field. Our review was based mainly on
results obtained in
papers~\citep{Astashenok2015,Astashenok2017,Astashenok2020}. The
mass--radius relation, mass profile and dependence of mass on
central density are given in comparison with general relativity.
From~a methodological point of view, we used metrics in
isotropical coordinates for the deriving of modified Einstein
equations. In~these coordinates, equations take relatively simple
forms. Inclusion of $R^2$-term leads to additional equation for
scalar~curvature.

For illustration we consider two equations of state {(EoS)} for
nuclear matter - GM1 \citep{GM1} and APR~\citep{APR}.

\section{Basic Equations for Non-Rotating Stars in Isotropic~Coordinates}

Einstein equations from general relativity have the following form
(in the natural system of units with $G=c=1$):
\begin{equation}
R_{\mu\nu}-\frac{1}{2}g_{\mu\nu}R={8\pi} T_{\mu\nu},
\end{equation}
where $R_{\mu\nu}$ are components of the Ricci tensor,
$R=g^{\mu\nu}R_{\mu\nu}$ is the scalar curvature and $T_{\mu\nu}$
is the energy-momentum tensor of~matter.

In a case of $f(R)$ gravity ({{first, second and third derivatives of $f(R)$ on $R$ should exist})} 
 with the~action
\begin{equation}
S=\frac{1}{16\pi}\int f(R) \sqrt{-g} d^{4} x
\end{equation}
equations became more complex
\begin{equation}\label{EinFR}
f_{R} R_{\mu\nu}-\frac{f}{2} \, g_{\mu\nu} -\left( \nabla_{\mu}
\nabla_{\nu}- g_{\mu\nu}\Box\right) f_{R}=8\pi T_{\nu\mu}.
\end{equation}
Here and further on, we omit argument $R$ for $f(R)$.
The~covariant D'Alambertian $\Box=\nabla^{\mu}\nabla_{\mu}$ is
introduced and $f_{R}$ simply means $df/dR$.

One can rewrite (\ref{EinFR}) in equivalent form
\begin{equation}\label{EinFR2}
f_{R}R_{\mu\nu}-\frac{1}{2}(f_{R}R-f)g_{\mu\nu}-\left(\frac{1}{2}\Box+\nabla_{\mu}
\nabla_{\nu}\right)f_{R}=8\pi
\left(T_{\mu\nu}-\frac{1}{2}g_{\mu\nu}T\right),
\end{equation}
where $T$ is the trace of energy-momentum tensor. It is useful to
make a foliation of spacetime by spacelike hypersurfaces
$\Sigma_{t}$ with unit vector $n_{\alpha}$ orthogonal to them.
Therefore, components of metric $\gamma_{\alpha\beta}$ induced on
$\Sigma_{t}$ are
\begin{equation}
\gamma_{\alpha\beta}=g_{\alpha\beta}+n_{\alpha}n_{\beta}.
\end{equation}
We consider the case of the non-rotating stars for which the
metric components do not depend on~time:
\begin{equation}
ds^{2}=-N^{2}dt^{2}+\gamma_{ij}dx^{i}dx_{j},
\end{equation}
Here $N$ is lapse function. Then one needs to project
Equations~(\ref{EinFR2}) twice onto hypersurface $\Sigma_{t}$,
twice along to normal vector $\vv{n}$ and once along $\vv{n}$ and
$\Sigma_{t}$. As a~result, one obtains three equations:
\begin{equation} \label{PDE1}
f_{R}\left(D_i D_j N - N\left\{ {}^3 R_{ij} + K K_{ij} -2 K_{ik}
K^k_{\ \, j}\right\}\right)=
\end{equation}
$$
=4\pi N \left[ (\sigma-\epsilon) \gamma_{ij} - 2 \sigma_{ij}
\right]-\frac{1}{2}(f_{R}R-f)N\gamma_{ij}-N\left(\frac{1}{2}\gamma_{ij}\Box
+D_{i}D_{j}\right)f_{R},
$$
\begin{equation}
f_{R}({}^3 R + K^2 - K_{ij} K^{ij}) = 16\pi
\epsilon+f_{R}R-f+2D^{i} D_{i} f_{R},
  \label{PDE2}
\end{equation}
\begin{equation}
f_{R}( D_j K^j_{\ \, i} - D_i K) = 8\pi  p_i
-n^{\mu}\nabla_{\mu}(D_{i}f_{R}), \label{PDE3}
\end{equation}
In these equations $K_{ij}$ are components of the tensor of
extrinsic curvature and $K=K^{i}_{i}$. The~3-dimensional covariant
derivatives $D_i$ are defined via 3-dimensional Christoffel
symbols ${}^3 \Gamma^i_{\ \,jk}$: \be
   D_i D_j N = \dder{N}{x^i}{x^j} - {}^3 \Gamma^k_{\ \, ij}
     \der{N}{x^k} ,
\ee \be D_j K^j_{\ \, i} = \der{K^j_{\ \, i}}{x^j}
     + {}^3 \Gamma^j_{\ \, jk} K^k_{\ \, i}
    - {}^3 \Gamma^k_{\ \, ji} K^j_{\ \, k} ,
\ee \be
   D_i K = \der{K}{x^i} .
\ee The components of the 3-dimensional Ricci tensor $^{3}R_{ij}$
and scalar curvature can be calculated via ${}^3 \Gamma^i_{\
\,jk}$ and its partial derivatives from standard~relations.

On the right-hand sides of Equations~(\ref{PDE1})--(\ref{PDE3}),
quantities $\epsilon$, $\sigma_{ij}$ and $p_{i}$ are defined by
relations from energy-momentum tensor:
\begin{equation*}
\epsilon=n^{\mu} n^{\nu} T_{\mu\nu},
\end{equation*}
\begin{equation}
\sigma_{ij}=\gamma^{\mu}_{i}\gamma^{\nu}_{j} T_{\mu\nu},\quad
\sigma=\sigma^{i}_{i}.
\end{equation}
\begin{equation*}
p_{i}=-n^{\mu}\gamma^{\nu}_{i}T_{\mu\nu}.
\end{equation*}
and have a sense of energy density, components of stress tensor
and a vector of energy flux density~correspondingly.

Then we take the trace of Equation~(\ref{PDE1}):
\begin{equation}
f_{R}D_{i}D^{i}N=Nf_{R}({}^{3}R+K^{2}-2K_{ik}K^{ik})+4\pi N
(\sigma-3\epsilon)+
\end{equation}
$$
+\frac{3}{2}N(f-f_{R}R)-\frac{3}{2}N\Box f_{R}-ND_{i}D^{i}f_{R}.
$$
From Equation~(\ref{PDE2}) it follows that
$$
f_{R}({}^{3}R+K^{2})=f_{R}K_{ij}K^{ij}+16\pi
\epsilon+f_{R}R-f+2D_{i}D^{i}f_{R}
$$
and therefore one can rewrite the previous equation as
\begin{equation}\label{PDE4}
f_{R}D_{i}D^{i}N=Nf_{R}K_{ij}K^{ij}+4\pi
N(\epsilon+\sigma)-\frac{1}{2}N(f_{R}R-f)-
\end{equation}
$$
-\frac{3}{2}N\Box f_{R}+ND_{i}D^{i}f_{R}.
$$

In the case of a non-rotating star, all metric functions depend
only on the radial coordinate. We use isotropic spatial
coordinates with a metric in the form
\begin{equation}\label{MetrSt}
ds^{2}=-N^{2}(r)dt^{2}+A^{2}(r)(dr^{2}+r^{2}d\Omega^{2}).
\end{equation}
For this metric, one can find that
$$
K=0,\quad K_{i j} K^{i j}=0.
$$
Three-dimensional scalar curvature is
\begin{equation}
   {}^3 R = -\frac{4}{A^2}
   \left( \triangle^{r}_{(3)}\ln A
   + \frac{1}{2} \left( \frac{d\ln A}{dr} \right) ^2 \right) \nonumber.
\end{equation}
Hereinafter, $\triangle^{r}_{(n)}$ means the radial part of
Laplace operator in n-dimensional euclidean space; i.e.,
$$
\triangle^{r}_{(n)}=\frac{d^{2}}{dr^2}+\frac{n-1}{r}\frac{d}{dr}.
$$
The energy-momentum tensor in the case of spherical symmetry is
simply $T_{\mu}^{nu}=\mbox{diag}(-\epsilon, p, p, p)$ where $p$ is
pressure of matter, and therefore
$$
\sigma^{r}_{r}=\sigma^{\theta}_{\theta}=\sigma^{\phi}_{\phi}=p,
\quad \sigma=3p.
$$
After algebraic calculations, the equations for metric functions
can be presented as (it is useful to introduce functions
$\eta=\ln(AN)$ and $\nu=\ln N$):
\begin{equation}\label{EQ1}
f_{R}\triangle^{r}_{(3)}\nu+\frac{1}{2}\triangle^{r}_{(3)}f_{R}=4\pi
A^{2}(\epsilon+3p)-\frac{A^{2}}{2}(f_{R}R-f)-f_{R}\frac{d\eta}{dr}\frac{d\nu}{dr}-
\end{equation}
$$
-\frac{1}{2}\frac{d\eta}{dr}\frac{df_{R}}{dr}-\frac{d\nu}{dr}\frac{df_{R}}{dr}.
$$
\begin{equation}\label{etaEQ}
f_{R}\triangle^{r}_{(4)}\eta+\triangle^{r}_{(4)}f_{R}=16\pi
A^{2}p-A^{2}(f_{R}R-f)-f_{R}\left(\frac{d
\eta}{dr}\right)^{2}-2\frac{d \eta}{dr}\frac{d f_{R}}{dr}
\end{equation}
For 4-dimensional scalar curvature, one can obtain an equation
from the trace of Einstein's equation:
\begin{equation}\label{CurvEQ}
\triangle^{r}_{(3)}f_{R}=\frac{8\pi}{3}
A^{2}(3p-\epsilon)-\frac{A^{2}}{3}(f_{R}R-2f)-\frac{d
\eta}{dr}\frac{d f_{R}}{dr}.
\end{equation}
Outside the star, the following conditions on  $\eta$, $\nu$ and
$R$ should be imposed:
$$
\nu\rightarrow 0, \quad \eta\rightarrow 0, \quad R\rightarrow
0\quad \mbox{for}\quad r\rightarrow\infty
$$
from the condition of asymptotical flatness on spatial infinity.
In general relativity, the solution of Einstein's equations
outside the star has the form:
\begin{equation}
A=\left(1+\frac{M}{2r}\right)^{2}, \quad
N=\left(1-\frac{M}{2r}\right)\left(1+\frac{M}{2r}\right)^{-1}.
\end{equation}
where parameter $M$ has sense of gravitational mass. Therefore,
the gravitational mass of a star can be defined from the
asymptotic behavior of $A$ at $r\rightarrow\infty$:
\begin{equation*}
M=2\lim_{r\rightarrow\infty}r(\sqrt{A}-1).
\end{equation*}
One should also take into account that the {{circumferential radius}} $\tilde{r}$ is 
\begin{equation*}
\tilde{r}=Ar.
\end{equation*}
Note that in the following the symbol "r" in the figures means
{{circumferential radial coordinate}.} Tilde is omitted
for~simplicity.

As illustrative example, we consider $R^2$ gravity for which
$$
f=R+\alpha R^2.
$$
Interesting results were obtained for $f(R)=R^{1+\epsilon}$
gravity in \citep{Capozz-1}. The~mass--radius relations in  metric
and torsional $R^2$ gravity were investigated in \citep{Capozz-3}.
For~a recent review of compact star models in modified theories of
gravity, see \citep{Olmo,Rev-2,Rev-3} and references~therein.

\begin{figure}[h!]
\centering
\includegraphics[scale=0.34]{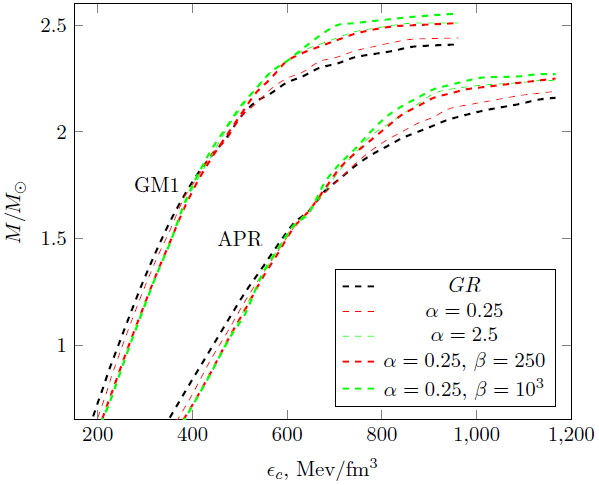}\includegraphics[scale=0.34]{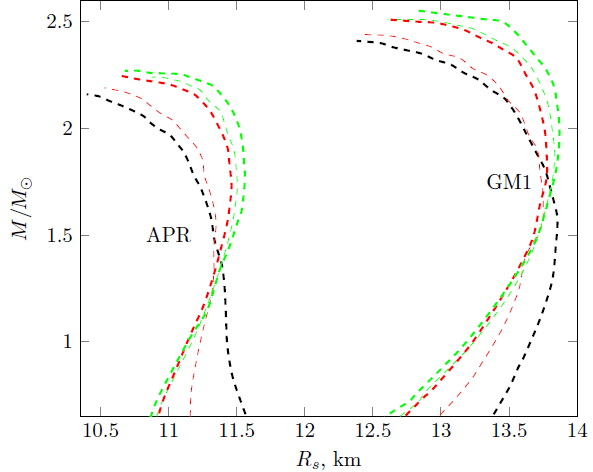}\\

\caption{Gravitational mass versus central density (left panel)
and radius (right panel) using equations of state (EoS) GM1 and
APR for some values of $\alpha$ and $\beta$ (in units of
$r_{g}^{2}$ where $r_g$ is gravitational radius of Sun) in
comparison with general relativity ($\alpha=0$). For~mass of axion
field, hereinafter, we take value $m_a=0.1$ in units of
$r_{g}^{-1}$.}\label{fig1}
\end{figure}

\begin{figure}[h!]
\centering
\includegraphics[scale=0.31]{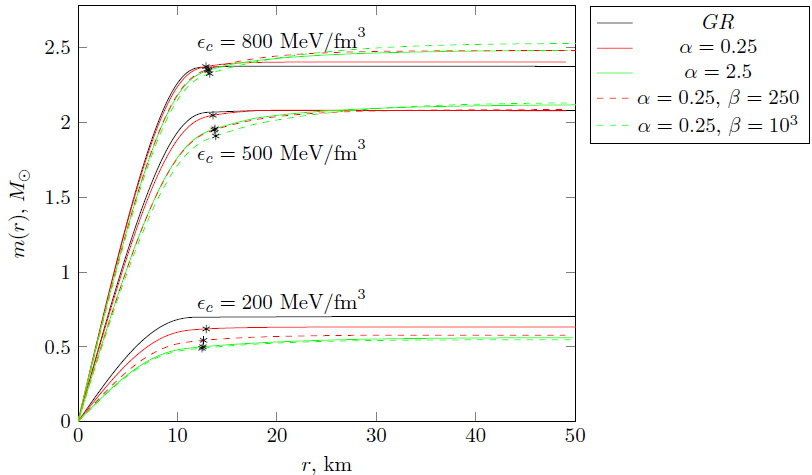}\includegraphics[scale=0.31]{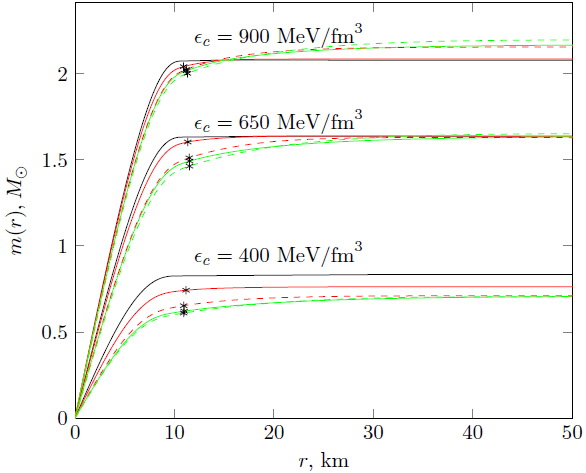}\\

\caption{Mass profiles for some central densities using EoS GM1
(left panel) and APR (right panel) for some values of $\alpha$ and
$\beta$ in comparison with general relativity. Asterisks
hereinafter designate points for which $\epsilon=p=0$ (surface of
star).}\label{fig2}
\end{figure}

\begin{table}[h!]
\centering \caption{Parameters of compact stars (masses and radii)
in general relativity ($\alpha=\beta=0$), simple $R^2$-gravity
($\beta=0$) and for  $R^2$-gravity with an axion field for some
values of central energy density using two EoS. In the last
column, the corresponding values of curvature in the center of
star are given.} \label{Table1}
\begin{tabular}{cccccc}
\toprule \boldmath{$\epsilon_{c}$}\textbf{,
MeV/fm}\boldmath{$^{3}$}  & \boldmath{\textbf{$\alpha$,
$r_{g}^{2}$}}   & \boldmath{\textbf{$\beta$, $r_{g}^{2}$}} &
\boldmath{\textbf{$M/M_{\odot}$}} &
\boldmath{\textbf{$R_{s}$, km}}      & \boldmath{\textbf{$R_{c}$, $r_{g}^{-2}$}}    \\
\hline
\multicolumn{6}{c}{GM1 EoS}\\
\hline
         & 0 & 0 & 0.70  & 13.42 & 0.048      \\
         & 0.25 & 0 & 0.63  & 12.95 & 0.034      \\
   200   & 2.5 & 0 & 0.56  & 12.52 & 0.0092      \\
        & 0.25 & 250 & 0.58  & 12.61 & 0.014     \\
        & 0.25  & 1000 & 0.55  & 12.46 & 0.0059      \\
         \hline
        & 0 & 0 & 2.07  & 13.55 & 0.049    \\
         & 0.25 & 0 & 2.07  & 13.58 & 0.050      \\
   500   & 2.5 & 0 & 2.11  & 13.79 & 0.019     \\
        & 0.25 & 250 & 2.08  & 13.73 & 0.016      \\
         & 0.25 & 1000 & 2.12  & 13.85 & 0.0068     \\
         \hline
        & 0 & 0 & 2.37  & 12.76 & $-$0.012      \\
        & 0.25 & 0 & 2.39  & 12.84 & 0.026     \\
   800  & 2.5 & 0 & 2.48  & 13.11 & 0.015     \\
         & 0.25 & 250 & 2.48  & 13.07 & 0.0092      \\
        & 0.25  & 1000 & 2.52  & 13.22 & 0.0057      \\
\hline
\multicolumn{6}{c}{APR EoS}\\
\hline
         & 0 & 0 & 0.83  & 11.47 & 0.087      \\
        & 0.25  & 0 & 0.76  & 11.17 & 0.058      \\
   400   & 2.5 & 0 & 0.72  & 10.95 & 0.014      \\
         & 0.25 & 250 & 0.71  & 10.95 & 0.018      \\
         & 0.25 & 1000& 0.71  & 10.90 & 0.0073      \\
         \hline
         & 0 & 0 & 1.63  & 11.30 & 0.067      \\
         & 0.25 & 0 & 1.63  & 11.33 & 0.066      \\
   650  & 2.5 & 0 & 1.63  & 11.50 & 0.021     \\
         & 0.25 & 250 & 1.64  & 11.45 & 0.018 \\
         & 0.25 & 1000 & 1.65  & 11.53 & 0.0076 \\
         \hline
        & 0  & 0 & 2.07  & 10.93 & $-$0.031      \\
        & 0.25 & 0 & 2.08  & 10.94 & 0.031      \\
   900 & 2.5 & 0 & 2.17  & 11.23 & 0.016     \\
       & 0.25  & 250 & 2.16  & 11.18 & 0.014     \\
       & 0.25  &1000 & 2.20  & 11.32 & 0.0062     \\
\hline
\end{tabular}

\end{table}

\begin{figure}[h!]
\centering
\includegraphics[scale=0.36]{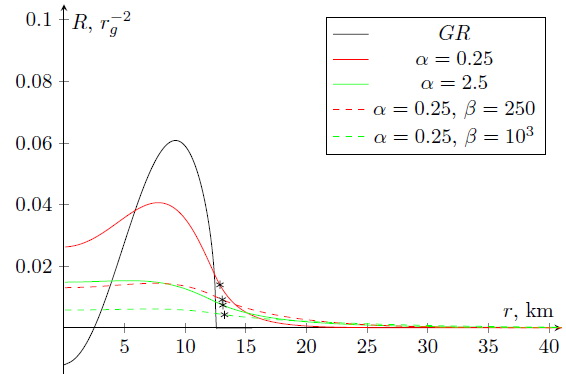}\includegraphics[scale=0.36]{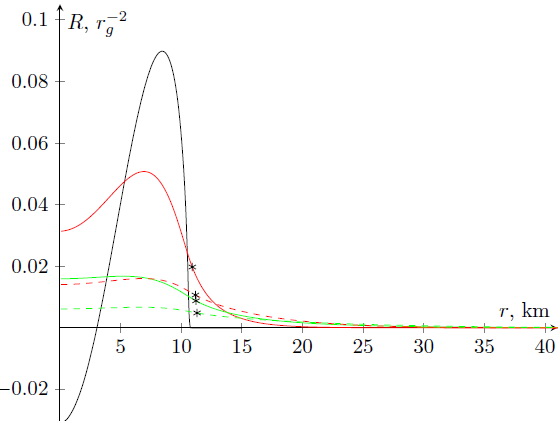}\\
\includegraphics[scale=0.36]{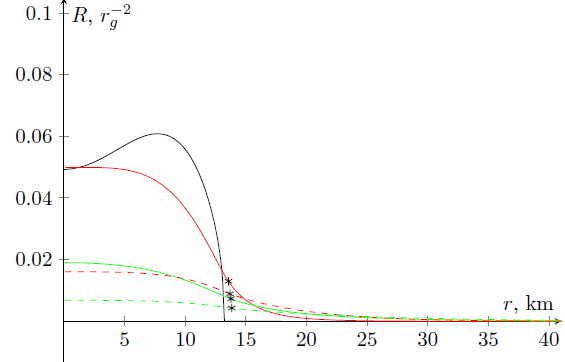}\includegraphics[scale=0.36]{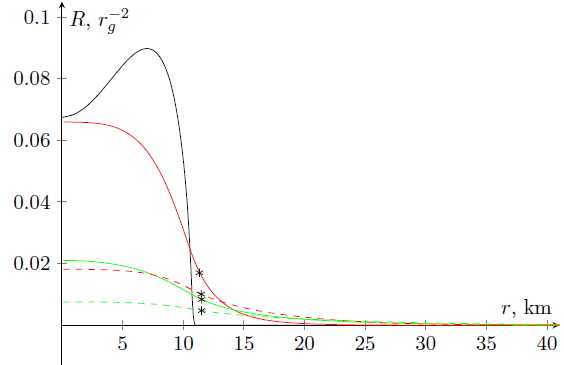}\\
\includegraphics[scale=0.36]{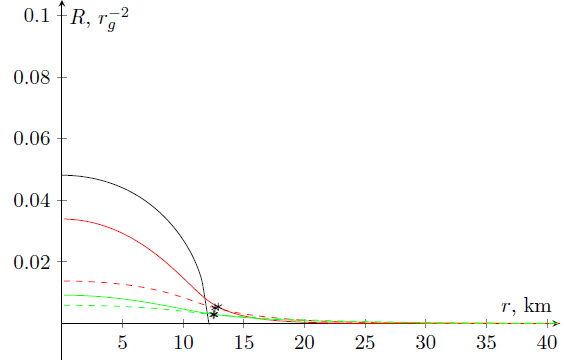}\includegraphics[scale=0.36]{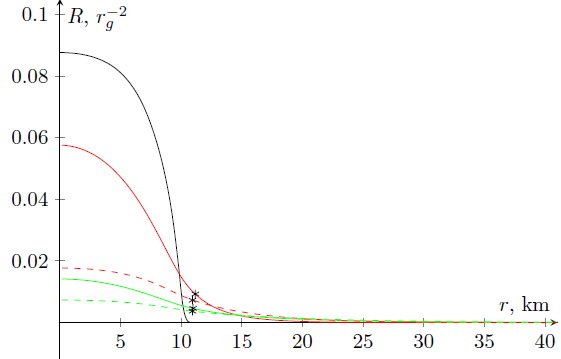}\\

\caption{Dependence of scalar curvature (in units of $r_{g}^{-2}$)
on radial coordinates using EoS GM1 (left panel) and APR (right
panel) for some values of central density and parameters $\alpha$
and $\beta$. For~GM1 EoS, $\epsilon_c$ is 800, 500 and 200
MeV/fm$^3$ (up to down), and for~APR, EoS---900, 650 and 400
MeV/fm$^3$ (up to down).}\label{fig3}
\end{figure}\unskip

\begin{figure}[h!]
\centering
\includegraphics[scale=0.36]{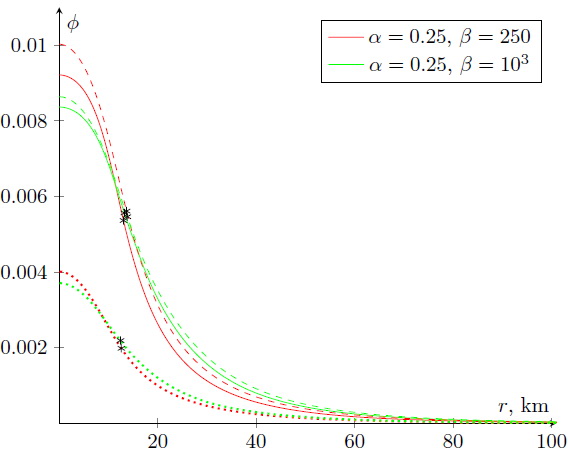}\includegraphics[scale=0.36]{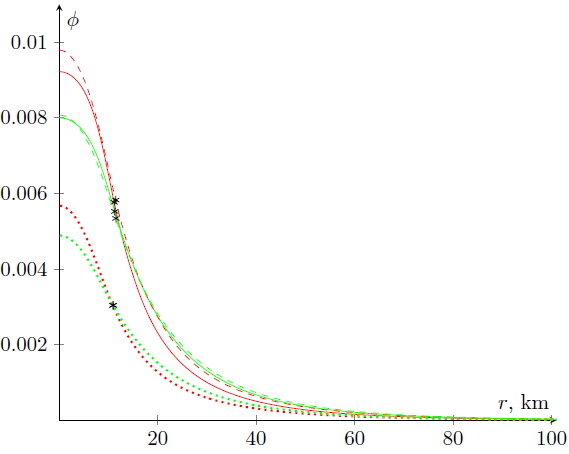}\\

\caption{Dependence of scalar field on the radial coordinates
using GM1 (left panel) and APR EoS (right panel) for some values
of central density and parameter $\beta$ in comparison with
general relativity for $\alpha=0.25$. Solid, dashed and dotted
lines correspond to $\epsilon_{c}=800\mbox{, }500\mbox{, }200$
MeV/fm$^3$ (GM1) and $\epsilon_{c}=900\mbox{, }650\mbox{, }400$
MeV/fm$^3$ (APR) correspondingly.}\label{fig4}
\end{figure}

\section{R-Square Gravity with a Scalar Axion~Field}

From results of simple $R^2$-gravity, it follows that
manifestations of modified gravity are observable only in cases of
large contribution from the $R^2$-term. The~addition of a scalar
field in the simple model with non-minimal interaction with
gravity allows one to construct solutions for which $R^2$-term
plays a significant role only inside~the star.

How can we motivate this extension of the simple model?
Astrophysical data about bullet cluster and {cluster MACSJ0025}
(see (\cite{Markevitch,Clowe,Robertson,Bradac}) speak in favor of
the particle nature of dark matter.
For a long time it was considered that dark matter was nothing other than so-called WIMPs 
 from models of supersymmetry in particle
physics. Unsuccessful experiments in WIMP searches (see for
example \citep{Ahmed,Davis,Davis2,Roszkowski,Schumann}) gave rise
to other hypotheses. In~particular, a realistic explanation is
that dark matter consists of axions
\citep{Sakharov,Sakharov-2,Sakharov-3,Marsh,Marsh-2,Oikonomou,Oikonomou-2,Oikonomou-3,Cicoli,Fukunaga,Caputo}.
In~contrast to failed experiments for WIMPs, there are some
indications in favor of the existence of axions (see
\citep{ADMX,Quellet,Safdi,Avignone,Caputo-2,Caputo-3,AX-1,Rozner}).
Axion emission can appears in the process of the cooling of
neutron stars \citep{Sedrakian}. Masses of axions can be very low
(theoretical estimations give value in the wide range $\sim
10^{-12}-10^{-3}$ eV). The~possibility of axions detection is
based on axion-photon interaction in the presence of magnetic
fields \citep{AX-3,AX-4,AX-5}.

The contribution to action from a free axion scalar field $\phi$
with mass $m_{a}$ is assumed in the following form
\begin{equation}\label{AXI}
S_{\phi}=\int
d^{4}x\sqrt{-g}\left(-\frac{1}{2}\partial^{\mu}\phi\partial_{\mu}\phi-
\frac{1}{2}m_{a}^{2}\phi^{2}\right).
\end{equation}
The solution for axion field $\phi$ with spherical symmetry can be
considered as a core with typical size $\sim m^{-1}$ (or $h / mc$
in SI units).  The~radius of a neutron star is 10--15 km;
therefore, the size of the axion core and the radius of the star
are comparable for $m_{a}\sim 10^{-11}$ eV. The~interaction term
in the Lagrangian equation is in simple form
$$
\mathcal{L}_{int}=\frac{\beta R^{2} \phi}{16\pi}
$$
In this case the equation for scalar field $\phi=\phi(r)$ is
written as
\begin{equation}\label{scalEQ}
\triangle^{r}_{(3)}\phi=A^{2}m_{a}^{2}\phi-\frac{A^{2}}{16\pi}\beta
R^2-\frac{d\phi}{dr}\frac{d\eta}{dr}.
\end{equation}

Therefore, one can consider the possibility of the existence of an
axion ``core'' containing dark matter in the center of the star.
The~contribution to energy density from such a core is negligible
itself, and therefore could not influence  the parameters of a
star. However, the~assumption of coupling $\sim R^2\phi$ can lead
to non-trivial deviations from general relativity.

For the case of function $f_{R}=f_{R}(R,\phi)$ depending also on
scalar field $\phi$, Equations~(\ref{EQ1})--(\ref{CurvEQ}) are
valid, but  one need only take into account that radial
derivatives of $f(R,\phi)$ are
\begin{equation*}
\frac{df_{R}}{dr}=f_{RR}\frac{dR}{dr}+f_{R\phi}\frac{d\phi}{dr},
\end{equation*}
\begin{equation*}
\frac{d^{2}f_{R}}{dr^2}=f_{RR}\frac{d^{2}R}{dr^2}+f_{RRR}\left(\frac{dR}{dr}\right)^{2}+f_{R\phi}\frac{d^{2}\phi}{dr^2}+f_{R\phi\phi}\left(\frac{d\phi}{dr}\right)^{2}+
2f_{RR\phi}\frac{dR}{dr}\frac{d\phi}{dr}.
\end{equation*}

The Equations~(\ref{EQ1})--(\ref{CurvEQ}) (and (\ref{scalEQ}) for
model with axion field) can be integrated with boundary conditions
at spatial infinity and given central density $\epsilon_c$.
The~surface of star corresponds to $\epsilon=p=0$. We use also the
consequence of Bernoulli theorem according to which for
non-rotational star
$$
H+\nu=\mbox{const},
$$
where $H$ is so called log-enthalpy
$H=\ln\left(\frac{\epsilon+p}{n_{b} m_{b}}\right)$. Here $n_b$ is
particle density and $m_b$ is mean baryon mass. Therefore, from
function $\nu(r)$ we can also define dependence of energy density
and pressure from radial coordinate in process of integration.
{{We use the self-consistent-field method for resolution of
equations}}
{(this method for rotating stars in general relativity is described in detail, for example, in \citep{Gourg})}. 

\section{Discussion of Results}

An illustration of the dependencies of the gravitational mass of a
star on radius and central density are given on Figure~\ref{fig1}
for model (\ref{AXI}) in comparison with general relativity and
simple $R^2$-gravity. We use two well-known equations of state
from nuclear physics; namely, GM1 (without hyperons) and APR.
From~Figure~\ref{fig1} one can see that for some value of central
density (for given EoS), masses and radii of star configurations
are very close to values in general relativity. Below~this
density, the radii and masses decrease in comparison with general
relativity. The opposite situation takes place for larger
densities. Deviation from general relativity is maximal for stars
with maximal masses for a given~EoS.

Additionally, it is interesting to consider the dependence of mass
confined by sphere with radius $r$ from $r$ (see
Figure~\ref{fig2}). In~general relativity,
$m(r)=M_{s}=\mbox{const}$ is constant outside the star surface.
But~in $R^2$ gravity and its extension with a scalar field, there
is a contribution to gravitational mass outside the surface of a
star. The~gravitational mass confined by the star's surface for
two models is always smaller than $M_{s}$. For~relatively large
central densities, the additional contribution to gravitational
mass overcomes this smallness, and the gravitational mass for a
distant observer~increases.


These features can be understood from the behavior of scalar curvature (Figure \ref{fig3}). Outside the star surface, there is an area in which scalar curvature is nonzero, in contrast with general relativity~\cite{Astashenok2017}. For~large $\alpha$ and for a model with axion field scalar curvature inside, the star slowly decreases the curvature. 
 Some results about masses and radii are given in Table \ref{Table1}. There is some equivalence between pure $R^2$-gravity and model (\ref{AXI}): for example, for $\alpha=0.25$ and $\beta=250$ we have similar results for mass as in the simple model for $\alpha=2.5$. The~explanation is simple. From~the dependence of the scalar field from radial coordinate (Figure~\ref{fig4}) one can estimate the contribution of the $\sim \beta \phi$ term into an effective value of $\tilde{\alpha}$:
$$
\tilde{\alpha}=\alpha + \beta\phi.
$$
The model with an axion field can be considered as an $R^2$-model
with non-constant parameter $\tilde{\alpha}$. If~mean value of
$\tilde{\alpha}$ is $\approx \alpha$, results for the mass and
radius of the star will be similar. The dependence of scalar field
$\phi(r)$ also demonstrates that the contribution of $\phi R^2$
grows with central~density.

One notes also, the weak dependence of mass increasing $\delta m$
from the value of parameter $\beta$. If~$\beta$ increases, the
mean value of curvature inside of star decreases, and then the
contribution of $\beta \phi^2 R$ grows not so rapidly as $\beta$.
In~principle there is some upper limit on $\delta m$ in this model
(as for in simple $R^2$-gravity) close to considered. One should
note that $\delta m$ is the same for GM1 and APR EoS for high
masses. For example, $\delta m\approx 0.15M_{\odot}$ in comparison
to general relativity for $\alpha=0.25$ and $\beta=10^3$.
The main question of course is the possibility of discrimination between these models and general relativity. Unfortunately, we have no well established mass--radius dependence for neutron stars from observations (only masses can be measured with high accuracy). {{One should mention recent papers}} 
 {\citep{Riley,Miller,NICER} in which the
authors considered limits on mass and radius for pulsar PSR
J0030+0451. Some EoS can be excluded due to these data, and for
example, APR EoS of course is under question in general
relativity. For~our model this statement is also valid, because
for intermediate mass, the possible value of the radius differs
from the GR value negligibly in comparison with the error of
measurements from NICER (\boldmath{$\sim$}1~km). For~GM1 EoS
satisfying these data, the picture is the same: because GR fits
these data well, our theory is also valid.} The second difficulty
is the uncertainty in the details of the equation of state for
dense matter. We hope that further progress in astronomical
observations and high energy physics can give an answer to this
question. Additionally, it is useful to consider EoS-independent
relations for neutron star properties~\citep{REL,REL-2}. We plan
to address this issue in future papers.


\end{document}